\documentclass[a4paper, 12pt]{article}
\usepackage[dvips]{graphics}
\begin{document}

\title{\huge Binary Particle Model of Weak Interactions}

\author{ F.  N.  Ndili \\
Physics Department \\
University of Houston, Houston, TX.77204, USA.}

\date{February 2011}

\maketitle

\begin{abstract}
We introduce the new concept of binary particle as the basic
matter unit that participates in weak interactions and not any one
fermion singly. We state the quantum numbers of this binary
particle,  and show the concept leads us to a natural explanation
of the standard model puzzle of the origin of flavor mixing and
the CKM matrix.  Certain other puzzles of the standard model such
as the absence of flavor changing neutral currents (FCNC), are
also explained naturally by the binary particle model. These
puzzles are currently thought to be esoteric properties of electro
weak interactions that have origins in physics beyond the standard
model at some ultra high energy scales. We show that this is not
necessarily the case.

\end{abstract}

{\it{Keywords: weak interactions, flavor mixing, binary particles}\/}\\
{\bf PACS: 14.80Bn }\\
E-mail: frank.ndili@gmail.com

\newpage

\section{Introduction}
One of the puzzles of the Standard model of electroweak
interactions  is quark flavor mixing, the extent of which  is
embodied in the Cabibbo-Kobayashi-Maskawa (CKM) matrix [1,2].
Similar flavor mixing occurs among leptons.  The standard model
has no fundamental explanation for this fermion flavor mixing.
Rather the standard model attributes the flavor mixing to an inter
play between so-called gauge eigenstates of fermions  wherein
fermions interact and couple gauge invariantly to mass giving
Higgs scalar fields, as against  a separate eigenstate of the same
fermions called mass eigenstate in which the fermion mass matrix
gained from Higgs vacuum expectation value is displayed diagonal.
Needless to say this explanation of flavor mixing as arising from
the rotation in flavor space from gauge eigenstates to mass
eigenstates, is not fundamental. It is tied to the Higgs model of mass. \\

Our question in this paper is can there be a more fundamental
origin  and explanation of this flavor mixing and the CKM matrix,
which does not necessarily rely on the specific Higgs model of
mass and Yukawa couplings, but more intrinsic to the nature of
weak interactions? We find we can gain such a more fundamental
perspective on CKM flavor mixing and other puzzles of the standard
electroweak model, by introducing the new concept of binary
particle, as the basic unit of fermionic matter that participates
in weak interactions, and actually defines what is weak
interaction.  We explain in section 2 this new concept of binary
particle and define its quantum numbers. Then in the remaining
sections, we show how binary particles throw  light on  these
current puzzles of the  standard model. \\

\section{The Binary particle and its quantum \\ numbers}

If we ignore the Yukawa-Higgs sector of the standard model, or any
one particular model of mass, one is left with a standard model
picture of weak interaction as a fundamental force characterized
by  V-A currents that couple to each other  through the
intermediary of heavy gauge bosons.  This means these currents
rather than the Yukawa-Higgs pieces, are the fundamental entities
that characterize weak interactions, along side the mediating
gauge bosons. Because the currents close on an $SU(2)_L $ Lie
algebra, weak interaction is also said to be  characterized by an
$SU(2)_L$ gauge symmetry, that leads to fermion fields
participating in weak interactions  being classifiable into
representations of $SU(2)_L$. It is an intrinsic observational
feature of weak interactions that only doublet representations are
realized. Therefore, for as long as there is no limit to the
number and variety of fundamental fermions in nature,  their weak
interaction doublet grouping will necessarily proliferate, leading
to a feature called family replication  which is another form of
the flavor puzzle of the standard model. Our stand point in this
paper is that the flavor mixing phenomenon, the flavor family
replication phenomenon, the fermion doublets only structure,  are
all aspects of the same intrinsic property of weak interactions
which we try to explain using the concept of the
binary particle. \\

We regard the characteristic currents and doublets of weak
interactions as a clue to the true nature of weak interactions and
the  puzzles associated with flavor standard model. We interpret
the standard model currents of weak interaction or equivalently
the $SU(2)_L$  doublets of fermion fields appearing in weak
interaction Lagrangian, as implying that fermion matter fields,
participating in weak interactions do so only in pairs, which pair
we may denote generally now as one composite entity or particle B
= B(a,b), where $a$ and $b$ are the individual particle fermions
being paired, or naturally organized into partnership, to
participate as  entity B, in weak interactions. We assert that
there is no other way any matter field, fermion (or scalar), can
participate in weak interactions, except through such paired two
particle entity B. This partnership entity B can be thought of
variously as an $SU(2)_L$ doublet or equivalently as a weak
current, and written: \\

\begin{equation}\label{eq: ndili1}
 B = \left(
\begin{array}{c}
a \\ b
\end{array}
\right) =  \bar a\gamma_\mu (1 - \gamma_5) b
\end{equation}

This entity B has a number of definite quantum numbers that make
it perceivable as one physical observable particle, in much the
same way that one considers as observable, other composite objects
like hadrons or binary stars. \\

As one quantum number or intrinsic property of the binary particle
B, we note that since the entity B  is a doublet of $SU(2)_L$, its
two individual members necessarily carry different weak isotopic
spin charge $I_z = +1/2$ or $I_z -1/2$. We may then say that B as
an entity carries or contains within it, an isotopic charge
differential or gradient isotopic charge $\Delta I_z \ne 0$ in
general, defined as $\Delta I_z = I_z^i - I_z^f$. This $\Delta
I_z$ becomes one quantum number of our entity B. We show below
that this weak isotopic spin charge $\Delta I_z$ carried by B has
only integer values: $\Delta I_z = +1, -1, 0$. \\

Before that, we attribute also  electric charge $Q_B$ to our
entity B through its weak current form $\bar a \gamma_\mu (1 -
\gamma_5) b$. We define this electric charge carried by B as
follows. We first interpret a current $\bar a \gamma_\mu (1 -
\gamma_5) b$ as a transition $ b \rightarrow a$ and so define the
electric charge carried by such a current or entity B as: $Q_B =
Q_b + Q_{\bar a}$. Illustrative examples are:\\

(1) $\bar d \gamma_\mu (1 - \gamma_5) u$ means $u_L \rightarrow
d_L$. It has $Q_B = Q_u + Q_{\bar d} = 2/3 + (+1/3) = + 1$. Such
$Q_B = +1$ entity B or current, becomes an emitter of $W^+$ gauge
boson, or an absorber (attractor)  of $W^-$ gauge boson. It has
$\Delta I_z = I_z^i - I_z^f = I_z^u - I_z^d = 1/2 - (-1/2) = +1 $.\\

(2) $\bar u \gamma_\mu (1 - \gamma_5) d$ means $d_L \rightarrow
u_L$. It has $Q_B = Q_d - Q_u = -1/3 -2/3 = -1$. Such $Q_B = -1$
entity B or current, becomes an emitter of $W^-$ gauge boson, or
an absorber (attractor) of $W^+$ gauge boson.  It has $\Delta I_z
= I_z^d - I_z^u = -1 $. \\

(3) $\bar e\gamma_\mu (1- \gamma_5) \nu_e$ means $\nu_{eL}
\rightarrow e_L$. It has $Q_B = +1$ and $\Delta I_z = I_z^i -
I_z^f = +1/2 -(-1/2) = +1$ \\

(4) The case $Q_B = \Delta I_z = 0$, called  the weak neutral
current case is also represented by B. Explicitly this case is
typified by: $(e, \nu_e)_L = \bar e \gamma_\mu (1 -
\gamma_5)\tau_3 \nu_e = \bar e \gamma_\mu (1 - \gamma_5) e - \bar
\nu_e\gamma_\mu(1 -\gamma_5) \nu_e$ to which $W_{3\mu}^o$ couples. \\

We see in each case that the electric charge $Q_B$ carried by
entity B, occurs in integer units $Q_B = +1, -1, 0, $ and is also
equal to the weak isotopic charge differential or gradient charge
$\Delta I_z$ carried by B. This means the entity B carries a
particular form of electric charge, one that is always equal to
the isotopic spin charge differential or isotopic gradient charge
$\Delta I_z$ carried by B. This becomes a further quantum number
of our new entity B, namely that it carries a special charge $Q =
\Delta I_z = +1, -1, 0$ \\

We call this entity B, having  the above attributes and quantum
numbers, a binary particle. It is a structured partnership of two
particles (a,b) organized as a basic unit of matter to participate
in weak interactions, and it carries integer electric charge $Q_B$
that is always equal to its integer isotopic spin differential or
gradient charge $\Delta I_z$.  With integer charges, binary
particles can be considered as observable physical particles, each
consisting  of a pair of loosely held quarks or a pair of leptons,
The holding is by the gradient $\Delta I_z \ne 0$ force in
weak isotopic charge space or flavor space.   \\

That such a holding together is possible can be seen from the fact
that the gradient isotopic charge differential $\Delta I_z$
between the two particles inside the binary particle B, should
trigger some cascading force in isotopic charge space, amounting
to a new force we can identify as the weak force itself.  The
analogy is that of two heights in a gravitational field from which
one ball can have a free fall towards the other. In the weak
interaction case, the free fall can be in either direction in
isotopic spin flavor space, accompanied by emission or absorption
of $W^\pm$ gauge bosons.  We state this as a third property or
attribute of our binary particle or weak binary formation, that
any one member of the doublet can have a "free fall" towards the
other, amounting to transition of one member into the other,
accompanied by emission or absorption of some weak gauge boson. \\

We assert finally that the gradient isotopic spin charge in its
totality of $Q_B = \Delta I_Z$ carried by a binary particle,  is
the basic charge source of all weak interactions, and  a direct
emitter and absorber of the weak gauge boson quanta $W^\pm,
W_3^o$. This inherent gradient character of the weak force can
produce an effect perceivable as one particle of the doublet or
binary, decaying or converting into the other member of the
doublet or binary, accompanied by emission or absorption of a
$W_\mu$ gauge boson, which was the early concept of weak
interaction as a single particle decay process. \\

What the binary particle model is now asserting is that there is
more to weak interactions than the old single particle decay
process. Weak interaction is never a single fermion affair. Rather
the binary particle B with its intrinsic two isotopic spin charge
levels and a differential $\Delta I_z \ne 0$, is a pre-requisite
basic matter unit for any weak interaction. This central point of
the binary particle model can be re-stated further to mean that it
is not the individual particle values of Q and $I_z$ that
determine weak interaction. Rather it is how two particles stand
in relation to each other in weak (flavor) isotopic spin charge
space, that determines whether a weak interaction or partnership
for weak interaction is possible or not, between them. The two
level weak isotopic charge system  which is another way of viewing
the binary particle B, is a pre-requisite mode in which the two
participating particles that make up B,  must first constitute
themselves before
they can experience or participate in weak interactions. \\

Even when one individual particle like $\mu \rightarrow e^- + \bar
\nu_e + \nu_\mu $ or $ d \rightarrow u + e^- + \bar \nu_e$ appears
to initiate alone the weak interaction process, our binary model
is asserting that the underlying weak driving force comes from the
gradient weak force $\Delta I_z \ne 0$ contributed jointly by the
two particles $(\mu, \nu_\mu)$ or $(u, d)$ and not any one
particle alone.  We say no one individual particle can set up by
itself, the gradient force that we recognize as inherent in weak
interaction. Therefore no one particle alone can generate or
participate alone in weak interactions. It is only the binary
particle as one entity that can participate in weak interactions.
Whatever emission or absorption of weak gauge bosons  there is in
a weak interaction,  must be seen as an affair of the binary
particle system of two particles, not any one particle in
isolation.

\section{Partnership Probabilities for Binaries}
Having now introduced and sufficiently defined our concept of
binary particles for weak interactions, we will put the concept to
test in various observed aspects and features of weak interactions
especially aspects where the standard model has no basic
explanation. We show how the binary particle enables us to
understand these puzzles of the standard model weak interactions. \\

We take the binary particle as defined and ask a natural question.
What principles go into  binary partnership formation for weak
interactions? Suppose we take a given fixed particle $a$ as a
prospective member of a binary particle B(a,b) while  $b$ is the
other particle. Then it can happen that we have a pool of
particles from which to choose partner $b$. Denote this pool by
$b_i$, $i = 1, 2, 3, 4......$. If particle $a$ has isotopic spin
charge $I_z = +1/2$, then each of particles $b_i$ in the pool will
have isotopic spin charge $I_z = -1/2$, and conversely.  Now it is
either the given particle $a$ has a fixed preferred binary partner
$b_i$ for some fixed $i$, or all the potential partners $b_i , i =
1, 2, 3, 4,....$ are equally viable partners, subject perhaps  to
normal quantum laws of probability. Varieties of observed weak
interaction processes  as well as early Cabibbo current
phenomenology [1], immediately rule out the first option. That is,
no one physical particle has one preferred permanent binary
partner for weak interactions, but a variety of such partners. We
are then left with the possibility that the binary partnership
$(a,b_i)$ is determined by quantum probability, \\

We implement this probability option by asserting that the
different potential weak iso-doublet partners $b_i$ of $a$, are in
constant competition among themselves to form binaries with
particle $a$. We state that each potential partner $b_i$ has at
all times non-zero probability of being found partnering with $a$,
or put differently,  that particle $a$ itself will at all times be
found partnering not with just one single particle $b_i$, but with
a full mixture of all viable partners, each having a certain
degree of presence in the mixture, this degree being variable with
time.  This means that our binary B = (a,b), has another intrinsic
property, namely that for a given physical particle $a$ as member
of B, the other member b is in general not one single physical
particle at any time, but a collection or mixture of several
physical particles. The time variability of each component's
presence in the mixture, means further more, that the entire
mixture $b$ oscillates with time, either into itself $b
\leftrightarrow b$, or into a different composition: $b
\leftrightarrow b'$. These effects become further intrinsic
attributes of our binary particle. \\

To see that these intrinsic attributes of the binary particle can
be the fundamental origin and essence of standard model flavor
mixing, CKM matrix and observed neutrino oscillations, we
calculate explicitly, the above partnership probabilities for the
binaries. Thus take the case of three physical particles $b_1,
b_2, b_3$ in competition to form binary partnership with a given
physical particle $a$ for purposes of weak interaction.  We form
the competitors into one omnibus quantum state : $ b = x_1b_1 +
x_2b_2 + x_3b_3 $ where the $x_i$ represent probabilities that at
a given time, $b_i$ is the physical particle in partnership with
particle $a$.  The physical particle fields $b_i$ being all
fermions, we take the $x_i$ to be all complex. We normalize the
probabilities
by: $\sum_i|x_i|^2 = 1 $ \\

To incorporate the possibility that the mixture $b$ can oscillate
not only into itself $b \leftrightarrow b$, but into a totally
different composition $b \leftrightarrow b'$ we write our
probability equation more fully as: \\
\begin{eqnarray}
b^1 & = &  x_{11} b_1 + x_{12} b_2 + x_{13}b_3 \nonumber\\
b^2 & = &  x_{21} b_1 + x_{22} b_2 + x_{23}b_3 \nonumber\\
b^3 & = &  x_{31} b_1 + x_{32} b_2 + x_{33}b_3
\end{eqnarray}

or \\
\begin{equation}\label{eq: ndili3}
\left(
\begin{array}{c}
b^1 \\ b^2 \\ b^3
\end{array}
\right) = \left(
\begin{array}{ccc}
x_{11} & x_{12} & x_{13}\\
x_{21}  & x_{22} & x_{23}\\
x_{31} & x_{32} & x_{33}
\end{array}
\right)  \left(
\begin{array}{c}
b_1 \\ b_2 \\ b_3
\end{array}
\right)
\end{equation}

where $b^\alpha$, $\alpha = 1, 2, 3$ are three orthogonal mixture
states always coupled to particle $a$; the $b_i$, $i = 1, 2, 3$
are the individual physical particles in constant competition to
couple in partnership with $a$. The individual particle
probabilities written now as $x_{ij}$  with their normalization
$\sum_i |x_{\alpha i}|^2 = 1$ condition realize a unitary matrix
$U U^\dagger = 1$, that appears to rotate the set of states
$b^\alpha$ into the set of competitor states $b_i$, and vice
versa. This matrix we recognize as standard model CKM  flavor
mixing matrix.  Thus  it can be said that the genesis of
unexplained standard model flavor mixing, is simply the inherent
competition among various candidate quarks, to form these binary
partnerships, in order to participate at all in weak interactions,
such binary partnership formation being a pre-requisite for any
matter particle to participate in weak interactions. \\

In $SU(2)_L$ doublet form,  the binary partnerships formed by
competing physical particles $b_i$, with the one particle a will
be written: \\
\begin{equation}\label{eq: ndili1A}
 \left(
\begin{array}{c}
a \\ b^1
\end{array}
\right)  =
 \left(
\begin{array}{c}
a \\ x_{11} b_1
\end{array}
\right)  +
 \left(
\begin{array}{c}
a \\ x_{12} b_2
\end{array}
\right)  +
 \left(
\begin{array}{c}
a \\ x_{13} b_3
\end{array}
\right)
\end{equation}

In the case of quarks we can take the competing  $b_i$ particles
of $I_z = -1/2$ to be $b_i = d, s, b$, while we take the $a$
particles in succession to be : $a = u, c, t$. Then our equations
become: \\
\begin{equation}\label{eq: ndili1B}
 \left(
\begin{array}{c}
u \\ b^1
\end{array}
\right)  =
 \left(
\begin{array}{c}
u \\ x_{11} d
\end{array}
\right)  +
 \left(
\begin{array}{c}
u \\ x_{12} s
\end{array}
\right)  +
 \left(
\begin{array}{c}
u \\ x_{13} b
\end{array}
\right)
\end{equation}

\begin{equation}\label{eq: ndili1C}
 \left(
\begin{array}{c}
c \\ b^2
\end{array}
\right)  =
 \left(
\begin{array}{c}
c \\ x_{21} d
\end{array}
\right)  +
 \left(
\begin{array}{c}
c \\ x_{22} s
\end{array}
\right)  +
 \left(
\begin{array}{c}
c \\ x_{23} b
\end{array}
\right)
\end{equation}

\begin{equation}\label{eq: ndili1D}
 \left(
\begin{array}{c}
t \\ b^3
\end{array}
\right)  =
 \left(
\begin{array}{c}
t \\ x_{31} d
\end{array}
\right)  +
 \left(
\begin{array}{c}
t \\ x_{32} s
\end{array}
\right)  +
 \left(
\begin{array}{c}
t \\ x_{33} b
\end{array}
\right)
\end{equation}

giving, relative to u,c,t as non-competing binary partnership
particles: \\
\begin{equation}\label{eq: ndili3A}
\left(
\begin{array}{c}
b^1 \\ b^2 \\ b^3
\end{array}
\right) = \left(
\begin{array}{c}
b_u \\ b_c \\ b_t
\end{array}
\right) = \left(
\begin{array}{c}
d" \\ s" \\ b"
\end{array}
\right) = \left(
\begin{array}{ccc}
x_{ud} & x_{us} & x_{ub}\\
x_{cd}  & x_{cs} & x_{cb}\\
x_{td} & x_{ts} & x_{tb}
\end{array}
\right) . \left(
\begin{array}{c}
d \\ s \\ b
\end{array}
\right)
\end{equation}
where $b_u$ is the composite competitor state of d,s,b, that is
always in contention when we consider how the individual particles
d,s,b, form partnerships with u. Similarly  $b_c$ is that
composite competitor state of d,s,b, that is in contention when we
consider partnership formations of d,s,b  with particle c. Then
$b_t$ relates to partnership formations of d,s,b with particle t.
\\

The 3 x 3 matrix $x_{ij}$ as deduced, is recognized as  the CKM
matrix and is seen to automatically satisfy the  binary
partnership probability normalization condition $\sum_i|x_{ia}|^2
= 1 $ for each row $a = u, c, t$ of the matrix.  This condition
for elements of the CKM matrix is called in the standard model,
unitarity condition. The condition presents a puzzle in the
standard model as the condition does not follow from any first
principles or standard model condition. The superiority therefore
of the binary particle model is seen here in the model obtaining
the unitarity property of the CKM matrix from
a basic probability principle of the binary particle model. \\

Also following readily from the binary particle model is the
standard model condition on elements of the CKM matrix, namely :\\

\[
\sum_{i=d,s.b}x_{ai}(x_{bi})^* = 0 ; a \ne b = u, c, t.
\]
This follows readily in the binary particle model as simple
orthogonality condition between the flavor mixture states
$b^\alpha$.  They satisfy:  $b^\alpha (b^\beta)^* = \delta^{\alpha
\beta} $ .

\section{Binary particles and the problem of  FCNC}

We consider next how the binary particle model illuminates the
standard model problem of flavor changing neutral current (FCNC).
It is known from several years of experimental studies that flavor
changing weak neutral currents  typified by $ (sd) = \bar s
\gamma_\mu (1 - \gamma_5)d ; (sb) = \bar b \gamma_\mu (1 -
\gamma_5)s, $ and such physical processes as $K^o \rightarrow
\mu^+ \mu^- ; K^o \rightarrow \pi^o \bar \nu \nu ; K^+ \rightarrow
\pi^+ \bar \nu \nu ; K^+ \rightarrow \pi^+ e^+ e^- ; B_d
\rightarrow X_s \bar l l$, do not exist, or if observed at all,
are greatly suppressed in a manner to suggest they occur only as
higher order processes and never a direct tree graph coupling of
these FCNCs with $Z_\mu^o$ gauge bosons.  The need to conform to
this strict experimental absence in nature of FCNC at tree level,
while entertaining it at higher loop order, led Glashow ,
Iliopoulos and Maini (GIM) in 1970[3], to predict the existence of
the charm quark (c) which was later discovered. They studied the
reaction $K^o \rightarrow \mu^+\mu^-$, and came also to the
conclusion that even when the charm quark c exists, it must enter
the weak interaction only in  partnership  with an iso- doublet
partner taken to be the Cabbibo quark $(s_\theta)$. \\

The standard model has no explanation for this strict exclusion of
tree level FCNC or direct coupling of weak gauge bosons to weak
neutral currents $(sd), (sb), (bd), (uc), (ut), (ct),$ except to
point to the CKM matrix as containing no such terms, but no
fundamental explanation even for the CKM matrix.\\

The binary particle model has a direct answer to the problem of no
tree level FCNCs.  The answer is that the two  particles within
each of the  pairs $(sd), (sb), (bd), (uc), (ut), (ct),$ that
feature in these FCNCs have the same isotopic spin quantum number
$I_z = - 1/2$ and therefore each pair has $\Delta I_z = 0$, and no
gradient weak force to drive the process. This means none of the
pairs is a binary partnership or binary particle that alone is the
entity that participates in weak interactions, being the carrier
of weak charge source $\Delta I_z$,  able to emit and absorb weak
gauge bosons.

\section{Binary particles and problem of fermion family replication}
Whether among quarks or among leptons, fermion doublets
proliferate, a puzzling feature of the standard model. The binary
particle model insight into this problem is that the binary
particle being the basic matter unit and carrier of weak
interaction charge $\Delta I_z$,  no one fermion or scalar can
participate in weak interaction except it first forms a doublet
with some partner. Therefore there must be several viable doublets
or binary particles occurring naturally.  Since binary particles
are physical particles like mesons and baryons, their numbers can
be limitless. Even with only six quarks and six leptons, the
number of pairings that realize binary particles is already large.
They can have a wide spectrum both in mass and other quantum
numbers, as long as each binary or doublet replication,  has the
basic defining property: $Q_B = \Delta I_z \ne 0$ in general.
Experiments should in fact see a wide  spectrum of binary
particles.

\section{Summary and Conclusion}
In summary, we have shown how useful the new concept of binary
particle in weak interactions can be in illuminating  some seeming
puzzles of the standard model of weak interactions. Observed
flavor mixing appears to be a case of quantum probability
competition among fermions to form binary particle partnerships
required for participation in weak interactions. What will be
interesting is to find parameters on which this probability
depends such as mass of competing fermions, such  that we can
calculate explicitly the various elements of the CKM
matrix. This problem is under study. \\

References: \\
1. N. Cabibbo, Phys. Rev. Lett. 10 (1963)531.

2. M. Kobayashi and L. Maskawa, Prog. Theor. Phys. 49 (1973) 652.

3. S. L. Glashow ,  J. Iliopoulos and L. Maiani,  Phys. Rev. D2
(1970) 1285.

\end{document}